# Entropy-Driven Preordering Assists Nucleation in Polyethylene

Renkuan Cao, Fan Peng, Yunhan Zhang, Hao Sun, Ziwei Liu, Tingyu Xu [*], and Liangbin Li[*]

*National Synchrotron Radiation Laboratory, Anhui Provincial Engineering Laboratory of Advanced Functional Polymer Film, CAS Key Laboratory of Soft Matter Chemistry, University of Science and Technology of China, Hefei 230026, China*

**For Table of Contents use only**

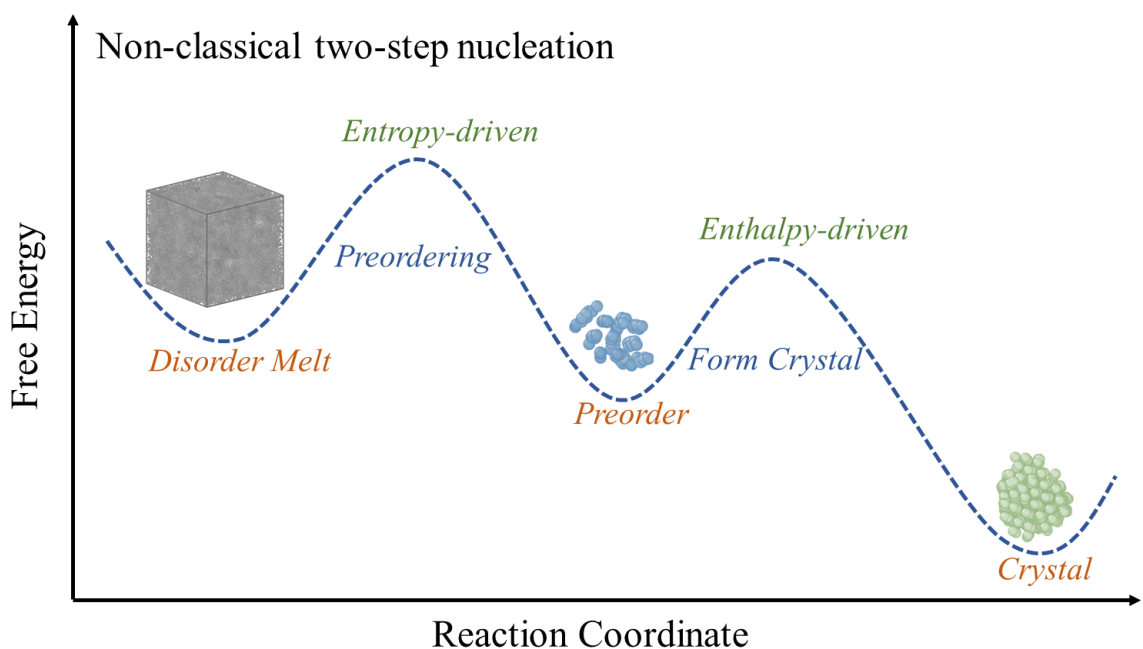

[*] Corresponding authors, E-mail: tyxu@ustc.edu.cn (T. Y. Xu); lbli@ustc.edu.cn (L. B. Li)

# Abstract

Non-classical two-step nucleation including preordering and crystal nucleation has been widely proposed to challenge the one-step nucleation framework in diverse materials, while what drives preordering has not been explicitly resolved yet. With molecular dynamics simulation, we find that two-step nucleation occurs in polyethylene, during which preordering precedes through the coupling between intrachain conformation and interchain orientation orders. Unexpectedly, preordering is driven by entropy rather than enthalpy, during which the interchain translational entropy gain compensates for the intrachain conformation entropy loss. This entropy-driven mechanism resolves the longstanding puzzle why flexible polymers with high entropy penalty still show high nucleation rate and opens a new perspective for understanding nucleation of synthetic and bio-polymers with conformation and orientation orders.

Crystallization is the most important first-order phase transition playing a crucial role in material processing, protein characterization, and pharmaceutical production.[1, 2] Hence, understanding the formation mechanism of crystals has always been attracting intense interest.[3, 4] According to the second law of thermodynamics, crystallization can be driven either by enthalpy[3, 5] or entropy.[6-8] In most material systems, the enthalpy gain caused by structural ordering is the main driving force of crystallization,[3, 9, 10] while for some special systems like colloid crystallization can be driven by entropy gain.[6, 8]

Either enthalpy- or entropy-driven crystallization mechanism is generally considered under the framework of classical one-step crystallization, while non-classical two-step crystallization has been observed in diverse systems such as colloid, mineral, polymer, protein and etc. with modern experimental[11-15] and simulation[16-22] techniques. The two-step crystallization models assume that preorder clusters with dense or structure-order form prior to the emergence of crystal, while crystal nuclei appear inside these clusters by structural ordering.[4] Compared to the one-step crystallization, preorder in two-step crystallization reduces the nucleation barrier and consequently accelerates crystallization,[4] as illustrated in Figure 1. Here enthalpy and entropy may play different roles in preordering and crystal nucleation, which, however, have not been clarified yet.

Due to the unique chain flexibility, polymer crystallization has to overcome large conformation entropy penalty caused by chain straightening, which is commonly regarded as an enthalpy-driven process.[23, 24] A longstanding puzzle in polymer community is why flexible polymers like polyethylene (PE) with large conformation entropy penalty still nucleate so fast that fully amorphous structure has never obtained by quenching. This puzzle may be answered by two-step nucleation. In preordering flexible chain transforms into straight segments through conformational ordering,[25] during which the enthalpy gain is weak while the conformational entropy loss $\Delta s_c$ is high. This implies that enthalpy gain is not the main driving force in preordering and an unrevealed driving force is required to balance conformational entropy loss $\Delta s_c$. Akin to liquid crystals, the orientation and conformational orders have been confirmed to occur before the onset of polymer crystallization.[18, 26-30] If the orientation order contributes a translational entropy gain $\Delta s_{tr}$ to serve as the hidden driving force in the preordering, contrary to common sense the largest barrier $\Delta s_c$ in polymer crystallization would not be overcome by enthalpy, but instead by entropy.

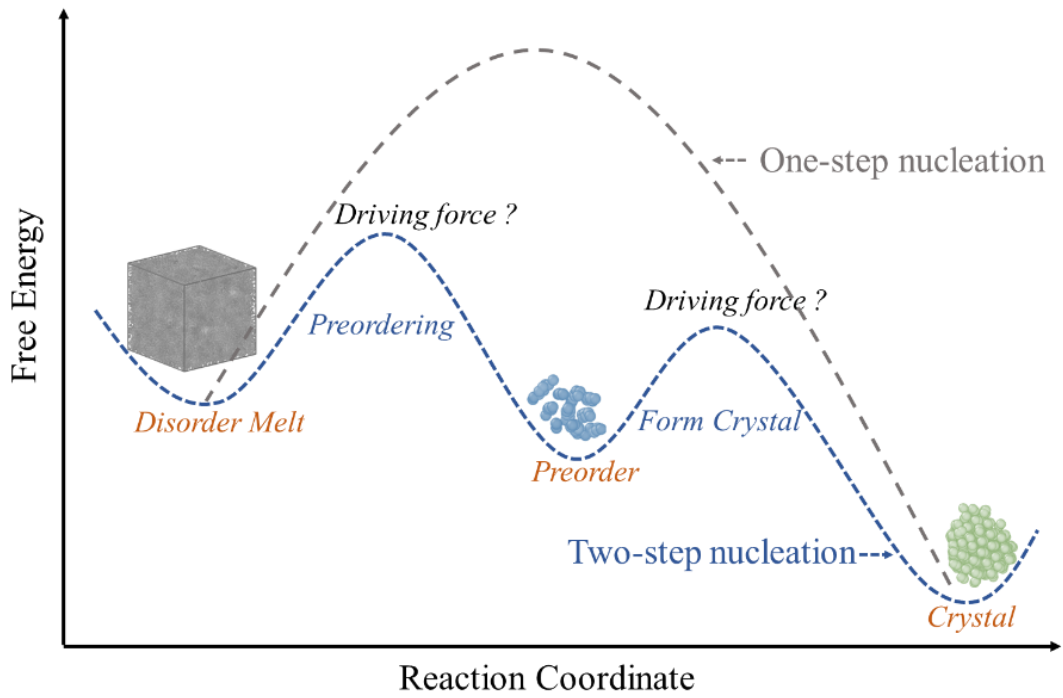

**Figure 1.** Schematic illustration of the free energy landscape for one-step versus two-step crystallization.

To unveil the respective roles of enthalpy and entropy in the preordering process of polymer crystallization, we employ molecular dynamics (MD) simulation to study the early stage of PE crystallization. Preorders with coupled intrachain conformation and interchain orientation orders are observed prior to crystallization, which is demonstrated to be an entropy-driven mechanism arising from excluded volume effects.

We utilize a united atom force field initially proposed by Paul, Yoon, and Smith,[31] and subsequently refined by Waheed (see Supporting Information).[32] All simulations are performed in the *NPT* ensemble, with the pressure maintained at 1 *atmosphere*. The pressure and temperature are controlled by the Nosé-Hoover barostat and thermostat, employing damping parameters of 5 *ps* and 0.5 *ps*. A time step of 5 *fs* is utilized throughout the simulations. Our system contains 250 C400 chains and follows the procedure illustrated in Figure 2A. The initial configuration undergoes an isothermal relaxation simulation at 600 *K* for 60 *ns* to achieve the equilibrium melt, see Supporting Information. Subsequently, the equilibrium melt is quenched to 300 *K* at a cooling rate of 60 *K/ns*, followed by an 800 *ns* isothermal simulation. The melting temperature is measured at 418.5 *K*, the undercooling is 28.3%, see Supporting Information.

By employing MD simulation, we study the early stage of PE quiescent crystallization (see Figure 2A), during which, the evolution of crystallinity ($\Phi_c$) and the average enthalpy change per particle ($\Delta h$) as a function of time are measured, see Figure 2B. Crystalline regions are identified using the cylindrical order parameter ($u_6 > 0.15$ and $u_8 > 0.15$), as proposed in our previous work.[19] The simulation system undergoes a re-equilibration of enthalpy in the first ~100 *ns* after quenching and an induction period of about 460 *ns* before the onset of crystallization, they are denoted as early stages. It is seen that $\Delta h$ almost presents a platform in the induction period. Crystallization begins at $t > 560$ ns, $\Delta h$ decreases gradually and the rate accelerates as the crystallization proceeds.

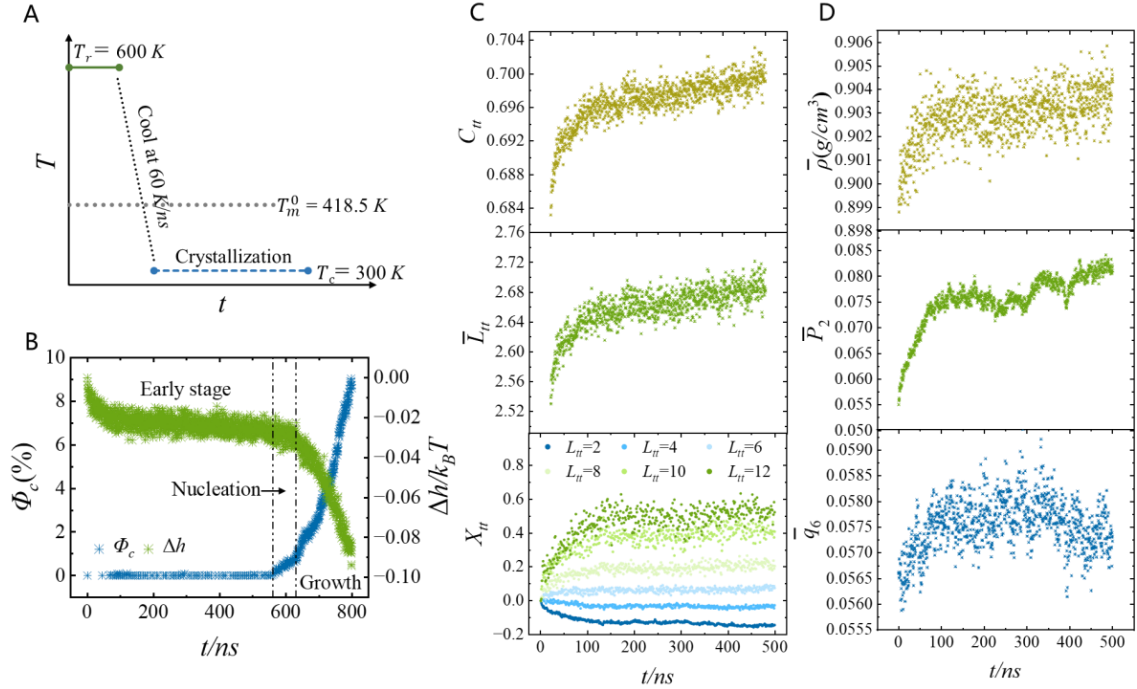

**Figure 2.** (A) Time-temperature protocols of the simulation procedure. (B) The crystallinity ($\Phi_c$) and the average enthalpy change per particle ($\Delta h$) versus time during isothermal crystallization. (C) The time evolution of intra-chain order including the *trans* fraction ($C_{tt}$), the average length of continuous *trans-trans* segments ($\bar{L}_{tt}$), and the rate of increase of *trans* segments, defined as $X_{tt} = [N_{L_{tt}}(t) - N_{L_{tt}}(0)]/N_{L_{tt}}(0)$. (D) The time evolution of three types of inter-chain order including average local density ($\bar{\rho}$), local orientation ($\bar{P}_2$), and bond orientation order parameter ($\bar{q}_6$).

The temporal variations of intra- and inter-chain order parameters in the induction period are illustrated in Figure 2C and 2D. It is characterized by the *trans* fraction ($C_{tt}$), the average length of continuous *trans-trans* segments ($\bar{L}_{tt}$), and the rate of increase of *trans* segments, defined as $X_{tt} = [N_{L_{tt}}(t)-N_{L_{tt}}(0)]/N_{L_{tt}}(0)$. Here, *gauche* and *trans* are defined as the conformation which dihedral angle $|\varphi| < 120°$ and $|\varphi| > 120°$, see Supporting Information. $N_{L_{tt}}(t)$ and $N_{L_{tt}}(0)$ denote the number of $L_{tt}$ segments at $t$ *ns* and 0 *ns*, respectively. As depicted in Figure 2C, both the $C_{tt}$ and $\bar{L}_{tt}$ initially increase steeply in the first 100 ns and then slowly over time. For short *trans* segments with $L_{tt} \leq 4$, $X_{tt}$ decreases with time. Conversely, for long *trans* segments with $L_{tt} > 4$, $X_{tt}$ increases with time. This indicates that the intra-chain conformational order (*gauche*-to-*trans* transition) occurs prior to the formation of the crystal. The time evolutions of three inter-chain order parameters, namely, average local density ($\bar{\rho}$), local orientation ($\bar{P}_2$), and bond orientation order parameter ($\bar{q}_6$) are systematically measured (see Figure 2D). These parameters are ensemble-averaged for each system snapshot (detailed calculation methods are elaborated in the Supporting Information). Similar to intra-chain parameters $C_{tt}$ and $\bar{L}_{tt}$, the inter-chain parameters $\bar{\rho}$, and $\bar{P}_2$ also exhibit rapid and gradual two-stage increases, while $\bar{q}_6$ follows a rapid increase and gradual decrease trend over time. These observations suggest that the inter-chain order, characterized by $\bar{\rho}$, $\bar{P}_2$, and $\bar{q}_6$, precedes the formation of the crystal. This observation is consistent with findings from previous experiments[30] and simulation results.[18, 27]

To verify the coupling between intra- and inter-chain order, contour maps of $\rho$, $P_2$, and $q_6$ in $t$-$L_{tt}$ space are plotted in Figure 3A. As $L_{tt}$ increases, $\rho$, $P_2$, and $q_6$ increase. Regions with higher $L_{tt}$ values correspond to higher values of $\rho$, $P_2$, and $q_6$, indicating a coupling between $L_{tt}$ and these inter-chain order parameters. To find the correlation relationship, we measure the Pearson correlation coefficient ($C(L_{tt}, O)$) of $L_{tt}$ and $O$ as a function of time $t$, where $O$ represents the $\rho$, $P_2$, and $q_6$. The $C(L_{tt}, O)$ is calculated by

$$C(L_{tt}, O) = \frac{\sigma[L_{tt}(t), O(t)]}{\sigma[L_{tt}(t)]\,\sigma[O(t)]} \tag{1}$$

where $\sigma[x,y]$ denotes the covariance between $x$ and $y$, and $\sigma[x]$ denotes the square root of variance of $x$. As shown in Figure 3B, $C(L_{tt}, P_2)$ increases over time and is consistently much larger than $C(L_{tt}, \rho)$ and $C(L_{tt}, q_6)$. It indicates that the coupling of conformation and density considered by Olmsted *et al.*[33] is extremely weak and insufficient to trigger spinodal decomposition. The predominant coupling mode between intra- and inter-chain order is the interplay of conformation and orientation. By this predominant coupling mode, preorders with coupled intra-chain conformation and inter-chain orientation orders appear prior to the emergence of crystal and assist the subsequent crystallization, see Figure 3C.

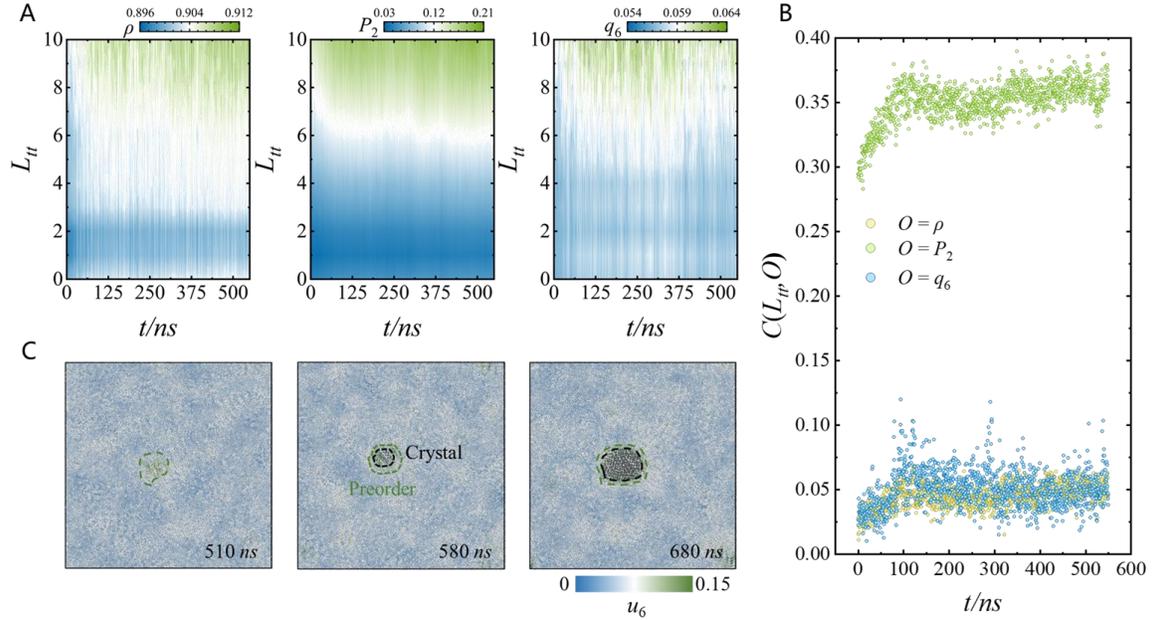

**Figure 3.** (A) The contour map of the length of continuous *trans-trans* segments $L_{tt}$ for crystallization time and inter-chain order parameters, including local density ($\rho$), local orientation ($P_2$), and bond orientation order parameters ($q_6$). (B) The Pearson correlation

coefficient ($C(L_{tt},O)$) of $L_{tt}$ and $O$ as a function of time, where, $O$ represents the $\rho$, $P_2$, and $q_6$. (C) System snapshot at 510 $ns$, 580 $ns$, and 680 $ns$. The amorphous particles and crystal particles are rendered with $u_6$ and black.

The enthalpy change in preordering is weak as shown in Figure 2B, suggesting an unrevealed driving force may exist. The *gauche-trans* transition involves a competition between the potential energy gain $\Delta h$ and the entropy loss $\Delta s$ per particle, which can be described by the free energy gap per particle $\Delta f = \Delta h - T\Delta s$. $\Delta f$ can be determined by counting the number distribution of *trans-trans* segments at different simulation times.[34] For a given time, the distribution of *trans-trans* segment numbers of different $L_{tt}$ follows the Boltzmann distribution, suggesting that their number is governed by energy. Thus, the formula $N_{L_{tt}} = N_0 \exp(-\Delta E/k_B T)$ is derived,[34] where $k_B$ is the Boltzmann constant and $N_0$ is a constant. By taking the natural logarithm, it can be transformed to

$$\ln N_{L_{tt}} = \ln N_0 + (-\frac{\Delta E}{k_B T}) \qquad (2)$$

In Figure 4A, the plot of $\ln N_{L_{tt}}$ versus $L_{tt}$ at 300 $ns$ shows a linear relationship (more data are plotted in the Supporting Information), leading to the assumption that $\Delta E$ is equal to $L_{tt} \Delta f$. Therefore, Eq. (2) can be rewritten as

$$\ln N_{L_{tt}} = \ln N_0 + (-\frac{L_{tt}\Delta f}{k_B T}) \qquad (3)$$

$\Delta f$ can be obtained from the slope of linear fit in Figure 4A. This linearity across all simulation times indicates that $\Delta f$ remains constant versus $L_{tt}$. Figure 4B depicts the variation of $\Delta f$ over simulation time (black circles). To facilitate the comparison, $\Delta f$ at 0 $ns$ is selected as the reference zero free energy. It is seen that $\Delta f$ decreases with time, suggesting that inter-chain coupling interaction (orientation) reduces $\Delta f$ and facilitates the *gauche-trans* transition.

$\Delta f$ can be divided into the intra-chain free energy ($\Delta f_{intra}$) along the chain direction and inter-chain coupling free energy ($\Delta f_{inter}$) deviating from the chain direction,

$$\Delta f = \Delta f_{intra} + \Delta f_{inter} = \Delta E_{intra} - T\Delta s_c + \Delta E_{inter} - T\Delta s_o \qquad (4)$$

$\Delta E_{intra}$ is the intra-chain energy difference between the *gauche* and *trans* state, which represents the change of intra-chain potential energy. It includes the energy gained from bond stretching, bond angle bending, and dihedral angle rotation. Until 550 *ns*, $\Delta E_{intra}$ has decreased by about 0.0122 $k_BT$, as shown in Figure 4B (blue circle). The change in intra-chain entropy is predominantly contributed by conformation entropy change $\Delta s_c$. In the present work, we quantify conformational entropy using dihedral angle entropy, as described in references.[35-37] It is written as

$$s_c = -k_B \left( \sum_{i=1}^{N_g} p_i \ln p_i + \ln N_g \right) \qquad (5)$$

where $p_i$ represents the probability of a dihedral angle occurring within the $i_{th}$ grid, with $N_g$ being the number of equally divided isometric grids spanning the entire range of dihedral angles. The variation of dihedral angle entropy ($\Delta s_c$) is insensitive to the value of $N_g$, as detailed in the Supporting Information. As depicted in Figure 4B (green circle), over the period up to 550 *ns*, $-T\Delta s_c$ increases by approximately 0.0235 $k_BT$. The *gauche-trans* transition increases chain rigidity, reducing freedom along the chain direction and thereby imposing an entropy penalty that hinders the formation of *trans* configurations. The change in the intra-chain free energy, $\Delta f_{intra} = \Delta E_{intra} - T\Delta s_c$, becomes positive over time, thereby inhibiting the *gauche-trans* transition. Evidently, the decrease in $\Delta f$ is driven by the change in inter-chain coupling free energy, $\Delta f_{inter} = \Delta E_{inter} - T\Delta s_o$. $\Delta E_{inter}$ represents the change in inter-chain potential energy (non-bond interaction energy) gain. As shown in Figure 4B (blue triangle), over the period up to 550 *ns*, $\Delta E_{inter}$ decreases by about 0.0034 $k_BT$, as illustrated by the green diamond. This reduction is significantly smaller compared to the decrease in $\Delta f$. From these findings, it is inferred that the decrease in $\Delta f$ is predominantly influenced by the change in coupling entropy $-T\Delta s_o$, which is calculated with Eq. (4), as depicted in Figure 4B (blue triangle). Therefore, this observation leads to the conclusion that the preorder coupling mechanism between conformation and orientation is entropy-driven.

Referring to Onsager's lyotropic liquid crystal theory,[38] we attribute the coupling entropy $\Delta s_o$ to the translational entropy $\Delta s_{tr}$, which arises from excluding the volume effects. The polymer chain is regarded as a freely-jointed chain composed of the Kuhn segment with length $L$ and diameter $D$, $\Delta s_{tr}$ can be written as[38-40]

$$\Delta s_{tr} = \frac{1}{2} k_B \rho_{mol} \int dw \int dw' f(w) f(w') V_{exc}(w,w') \qquad (6)$$

where the $\rho_{mol} = N/V$ is the number density of Kuhn segments, with $N$ being the number of Kuhn segments and $V$ the volume of the simulated box, respectively. The function $f(w)$ denotes the orientational distribution function of the Kuhn segment. Due to the unavailability of the exact orientation vector $w$ for each Kuhn segment, the unit chord vector $e$ (the vector connecting two adjacent particles within the same chain) is used as a substitute. The selection of the unit chord vector does not significantly alter $\Delta s_{tr}$, as detailed in the Supporting Information. The $V_{exc}(w,w')$ represents the excluded volume between two Kuhn segments oriented along $w$ and $w'$. It can be expressed as

$$V_{exc}(\boldsymbol{w},\boldsymbol{w}') = 2L^2 D |\sin\theta_{\boldsymbol{w},\boldsymbol{w}'}| + 2\pi L D^2 + \frac{4}{3}\pi D^3 \tag{7}$$

where $\theta_{w,w'}$ denotes the angle between vectors $\boldsymbol{w}$ and $\boldsymbol{w}'$. The diameter $D$ can be calculated using the second virial coefficient as $B_2 = \frac{2}{3}\pi D^3$ [41-44] and expresses as

$$B_2 = -2\pi \int [\exp\left(-\frac{u(r)}{k_B T}\right) - 1] r^2 \, dr \tag{8}$$

where $u(r)$ is the non-bond interaction term of the force field (see the Supporting Information for more detailed information). The length of Kuhn segment $L$ can be calculated by $L = \frac{C_\infty n l^2}{R_{max}}$, where $C_\infty$ is Flory characteristic ratio and $R_{max}$ is the contour length. Here, $l = 0.153$ nm represents the length and number of each C-C bond, and $n = 399$ is the number of C-C bonds per chain. By combining Eq. (6) to eq. (8), $\Delta s_{tr}$ can be quantitatively calculated. Figure 4C illustrates the linear relationship between $\Delta s_o$ and $\Delta s_{tr}$ with slope of 0.92, indicating that the entropy increases in $\Delta s_o$ is mainly contributed by $\Delta s_{tr}$.

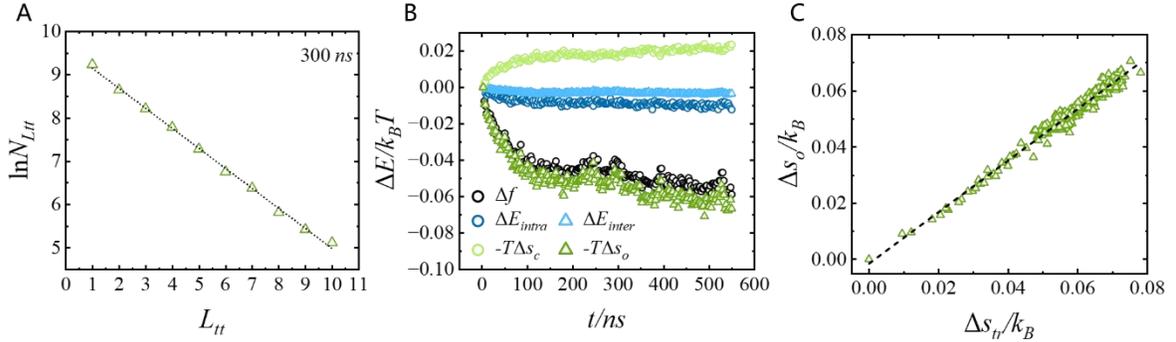

**Figure 4.** (A) The distribution of $\ln N_{L_{tt}}$ as a function of the length of continuous *trans-trans* segments $L_{tt}$ at 300 *ns*. The $N_{L_{tt}}$ represents the number of *trans-trans* segments. (B) The evolution of free energy and its components over simulation time. (C) Comparison of translation entropy with coupling entropy.

The existence and the nature of preorders have been widely reported in various material systems. Preorder reduces the nucleation barrier of crystal and accelerates crystallization is emphasized as compared to the classical one-step nucleation (see Figure 1). Whilst how preorder crosses the free energy barrier and what is the main driving force are not explicitly studied yet except in colloid and hard particle systems with designed entropy and enthalpy interactions. Our results reveal that the preordering of polymer precedes via coupling between conformation and orientation order. The translational entropy caused by the excluded volume effect of conformationally ordered rigid segments compensates for intrachain conformational entropy loss and drives preordering. Crystallization of flexible polymer is generally considered to be hard as it has to overcome the large conformation entropy penalty. This entropy-driven mechanism gives an answer to the longstanding puzzle that the crystallization rate of flexible polymers like PE is so fast that a fully amorphous structure has never been obtained. Through coupling between intrachain conformation and interchain orientation orders, the large conformation entropy penalty is compensated by the translational entropy, which results in the formation of preorder and subsequently accelerates the crystallization of flexible polymers. Current entropy-driven mechanism for preordering provides a generic model with two mutually compensating entropies, it may not only exist in synthetic polymers, but also presents in natural macromolecules like protein with similar conformation and orientation orders, for which non-classical two-step crystallization has been largely reported.

In summary, we reveal a two-step crystallization process of disordered melt → preorder → crystal. The first step is a process of mutual coupling between intrachain conformation and interchain orientation orders. The translational entropy gain caused by orientation order compensates for the conformation entropy loss. It is crucial, particularly for flexible polymers with weak intra-chain interactions, where a high entropy penalty hinders long *trans-trans* segment formation. For the second step (i.e., precursor → crystal), enthalpy gain is significant, indicating that it is enthalpy-driven. Our results reveal that enthalpy and entropy play different roles in these two steps, opening a new perspective to understanding polymer crystallization.

## ACKNOWLEDGMENTS

This work is supported by the National Key R&D Program of China (2020YFA0405800) and the National Natural Science Foundation of China (51633009).